Wealth and Poverty: The Effect of Poverty on Communities
Merrick Wang
Advisor: Dr. Robert Johnston, Associate Professor Emeritus of Finance at George Mason
University
January 21, 2018

**Abstract**

This paper analyzes the differences in poverty in high wealth communities and low wealth communities. We first discuss methods of measuring poverty and analyze the causes of individual poverty and poverty in the Bay Area. Three cases are considered regarding relative poverty. The first two cases involve neighborhoods in the Bay Area while the third case evaluates two neighborhoods within the city of San Jose, CA. We find that low wealth communities have more crime, more teen births, and more cost-burdened renters because of high concentrations of temporary and seasonal workers, extensive regulations on greenhouse gas emissions, minimum wage laws, and limited housing supply. In the conclusion, we review past attempts to alleviate the effects of poverty and give suggestions on how future policy can be influenced to eventually create a future free of poverty.

Wealth and Poverty: The Effect of Poverty on Communities

## I. Introduction

The world today sees the growth and development of nations and cities at an unprecedented rate with technological innovations. However, while the nation as a whole may prosper, there exists a moderately hidden issue: roughly 40 million Americans live in poverty. Poverty is often passed on for generations due to poor conditions: less access to education, increased drug and alcohol use, higher levels of disease, and greater risks associated with crime. Not only does poverty affect the individuals in hardship, but also it spreads its effects, such as consequences on health, education, and crime, on the surrounding community.

Overall, nearly four in ten Californians are living in or near poverty (Bohn, Danielson, & Thorman, 2018). Less educated individuals have higher poverty rates, despite the fact that most poor families having at least one working adult. According to the government's statistics of population, living costs, and federal assistance, California is the poorest state in the United States. The Supplemental Poverty Measure estimates California's poverty rate to be 20.4%, or over 8 million Californians.

Poverty's effects usually bring up several questions. First, how can poverty be effectively measured so that its effects on a community can be accurately calculated? Second, how does the affluence of a given community affect poverty levels? Lastly, how can we combat the issue of poverty, and hope to eradicate it completely? These questions are complicated due to the fact that many cases of poverty are not recorded due to being homeless or being undocumented in America.

While there exists an extensive literature on poverty, our purpose of writing this paper is to fathom the differences between high wealth and low wealth communities so that, when future policy is enacted, a greater difference can be made in the eradication of poverty.

In this paper, we will better comprehend the situation of people living in poverty in the Silicon Valley by analyzing communities in Santa Clara County. In Section II, we will discuss the methodology of measuring poverty. In Section III, we will analyze causes of individual poverty and poverty in the Bay Area. Section IV will compare poverty in high wealth communities with poverty in low wealth communities. The paper will conclude with analysis of past attempts to alleviate the poverty crisis and speculate on potential future solutions.

## II. Overview of Methodology: Measuring Poverty

Throughout history, poverty has been classified as multidimensional, in that it is not just simply related to income. People in poverty may suffer from social exclusion, indecent work, and poor health conditions. Additionally, they have less control over their resources (health, insurance, income, water) than the upper class. The poor class is a natural result of our stratified societies; as a result, there will always be an upper class and a lower class (Titumir & Rahman, 2013).

The most popular methodology to measure poverty is through the monetary approach, which accounts for the cash value that the poor spend on all "necessary goods," assuming that the poor spend all of their money on food, water, and clothing (Titumir & Rahman, 2013). A threshold determines the status of individuals or households: if expenditures for a certain family on "necessary goods" are below the threshold, then they are considered to be in poverty. However, this method is limited since the poor may spend their money on many "non-necessary" items (alcohol, drugs, leisure); furthermore, the more a family makes, the more they are willing to spend, which results in unreliable fluctuations of their expenses, especially for those families with seasonal jobs.

Otherwise, there are three groups of classification: objective poverty, subjective poverty, and multi-dimensional deprivation (Spain Instituto, n.d.). Using the objective method focuses on direct measurement or observation of the poor. Using the subjective method focuses on the poor's own self-awareness or other's perceived image of the poor. Multi-dimensional poverty factors in not just the basic needs of all humans, but also some "non-essential" needs such as education or quality of housing.

One type of multi-dimensional deprivation analysis is the the participatory approach.  The participatory poverty assessment is defined as "a family of approaches and methods to enable… people to share, enhance, and analyze their knowledge of life and conditions, to plan and to act" (Chambers, 1994). The approach recognizes that monetary approaches are significant to the poor, but factors such as social exclusion, speechlessness, or uselessness to a community are applied when calculating poverty. By recognizing social factors, it is easier to identify barriers to express opinions, reasons individuals become more poor, and how gender and race take a part in poverty.

Congress measures poverty through the Official Poverty Measure (OPM), which was created in the 1960s during Lyndon Johnson's "War on Poverty." To determine whether individuals are in poverty, OPM compares individuals' pre-tax cash income to poverty thresholds that vary by family size and family members' ages. The poverty threshold is based on three times the cost of a minimum food diet in 1963.

In 2011, after decades of research to improve poverty measure, the U.S. Bureau of Labor Statistics and the U.S. Census Bureau released the Supplemental Poverty Measure (SPM). SPM assesses more factors than OPM does, including family resources and geographic variation. Additionally, SPM adds benefits that help families meet basic needs and subtracts necessary expenses to better assess each household's net family resources (Fox, 2017). Congress releases poverty estimates using both SPM and OPM annually.

To help policymakers in California determine the impact of welfare programs on those in need, in 2013, the Stanford University Center on Poverty and Inequality and the Public Policy Institute of California designed the California Poverty Measure (CPM). In order to better assess California's poverty, CPM builds on the general methodology of Congress's SPM: while SPM utilizes estimates from national household expenditures, CPM accounts for discrepancies from national estimates to provide a more realistic measure of poverty in California. Specifically, CPM focuses on an upward adjustment using the latest benefit and assistance enrollment statistics from California and a geographic adjustment of poverty thresholds at the county level to better reflect home ownership costs around the state, which vary widely (Bohn, Danielson, Levin, Mattingly, & Wimer, 2013).

Since SPM and CPM have the same general structure, there are two basic approaches to measuring poverty:

| | OPM | SPM | CPM |
|---|---|---|---|
| Description | Cash Resource Based | Accounts for Family Resources and Expenses not included in the Official Measure | 1. Adjusts for Survey Under-Reporting<br>2. Aimed at Producing County-Level Estimates<br>3. Adjusts for Geographic Differences in Housing Costs<br>4. Includes Food Stamps and other non-Cash Benefits as Resources Available to Poor Families |
| Release Year | 1960s | 2011 | 2013 |
| Family Resources | 1. Pre-Tax Cash Income<br>2. Cash-Based Government Benefits | 1. Cash Income<br>2. All Government Benefits | |
| Expenses | N/A | -Taxes<br>-Out-of-Pocket Expenses for Work Expenses (commuting, child care)<br>-Out-of-Pocket Medical Costs | |
| Used in Assessment of Government Assistance | Yes | No | |

*Note.* Data from Danielson (2014) and U.S. Census (2017).

**III. Causes of Individual Poverty**

Medical accidents, layoffs, or defaults may send individuals into poverty. In a broader view, individuals' poverty may come from political instability, social inequality, or discrimination. However, in the Silicon Valley specifically, three additional major factors contribute to a higher rate of poverty: greenhouse gas regulations, high housing prices, and increases in the minimum wage.

Extensive regulations on greenhouse gas emissions make energy more expensive for individuals in the Silicon Valley. Individuals in California pay around 50% more than the national average for their electricity bills (Penn & Menezes, 2017). In a a 2015 Manhattan Institute study, Jonathan A. Lesser of Continental Economics concluded that "in 2012, nearly 1 million California households faced … energy expenditures exceeding 10% of household income. In certain California counties, the rate of energy poverty was as high as 15% of all households" (Jackson, 2018). Indeed, California produces so much solar energy at high costs that gifts of free solar power were given to Arizona for several weeks in early 2017 (Penn, 2017).

In 2015, more than four out of ten families in California have spent more than 30% of their total income on housing (Jackson, 2018). With high demand for employment opportunities (e.g. Google, Facebook), top-ranked schools, and excellent weather, the price of homes increases greatly. Furthermore, there is not much land available in the Bay Area for living - almost half of the Bay Area is preservation land or mountains. With policies such as the California Environmental Quality Act (CEQA) that increase the cost of completing a homebuilding project, housing supply becomes more limited, driving prices of homes higher. Moreover, the CEQA delays projects for homeless shelters, senior housing, and affordable housing for low-income residents (Collins, 2016).

Renters and homeowners with mortgages in the Silicon Valley are particularly likely to have unaffordable housing costs. According to the US Department of Housing and Urban Development, for housing to be considered affordable, costs should not exceed 30 percent of total household income. In 2015, 54.2% of renters were cost-burdened with their dwelling costs exceeding 30% of household income; of these cost-burdened renters, 53.1% were severely cost-burdened with their dwelling costs exceeding 50% of household income (Kimberlin, 2017). For individuals with lower incomes (less than 200% of the FPL[1]), 80.6% were cost-burdened for shelter costs (Kimberlin, 2017).

A thought experiment was conducted by the Public Policy Institute of California to determine the significance of housing prices on poverty. Bohn, Danielson, Levin, Mattingly, & Wimer (2013) posed the question, "What would poverty rates in California be if everyone in the state were to experience the housing costs of a low-cost county?" They concluded that California's overall poverty rate would drop by 7% and that child poverty rate would be more than 9% lower. As many Californians do not own their homes, housing increases the cost of living significantly.

In 2016, the governor of California signed a bill that increases the minimum wage in increments set to reach $15 by 2022. A Harvard University study by Luca & Luca (2017) found that "higher minimum wages increase overall exit rates for restaurants." Additionally, while 5-star restaurants had relatively no response to the minimum wage spike, " a one dollar increase in the minimum wage leads to a 14 percent increase in the likelihood of exit for a 3.5-star restaurant (which is the median rating)" (Luca & Luca, 2017). Entry-level workers, for whom the bill was designed to help, have their jobs threatened by the increase in minimum wage. When the first increase—from

---

[1] FPL stands for Federal Poverty Level. This threshold varies by the number of people in a household. For a two person household in 2018, the poverty guideline is $16,460.

$10.50 to $11.00—occurred, dozens of businesses began to lay off workers or leave the state (Saltsman, 2017). With a surplus of individuals willing to work but no businesses to hire due to the higher minimum wage, the workers become unemployed. The evidence indicates that minimum wage increases lead to a higher rate of poverty.

## A. Poverty in the Bay Area

Although the Bay Area[2] is one of the world's wealthiest regions, 11.3%, or 829,547 people, in the Bay Area were living in poverty in 2013 (Haveman & Massaro, 2015). Poverty rates in the Bay Area greatly increased during the Great Recession - in the early 2000s, OPM estimated Bay Area poverty at 9%, while during the Great Recession, OPM estimates increased to 12%.

Within the Bay Area, poverty rates vary widely, from a low of 7.8% in San Mateo County to a high of 13.8% in San Francisco. Although Bay Area poverty rates are lower than California's poverty rate, the lower rates may be misleading due to the significantly higher costs of living in the Bay Area. For instance, the cost of living in Solano County is significantly less than it is in San Francisco (Haveman & Massaro, 2015).

In the 2013 Census, the Official Poverty Measure estimated 191,805 individuals living in poverty in Santa Clara County. However, the figure is not accurate because the official poverty measure does not account for geographical differences - in parts of the Bay Area, cost of services is 6% higher, rent is 185% higher, and home prices are 250% higher than in the United States as a whole (Haveman & Massaro, 2015).

Haveman & Massaro (2015) compiled data for poverty in the Bay Area. The researchers found that in the Bay Area, African-Americans are most likely to be living in poverty (23.7%); white, non-Hispanic residents are least likely to be living in poverty (7.2%). Additionally, Haveman & Massaro (2015) concluded that the poverty rate for children and young adults (27.2%) is significantly higher than the general population's poverty rate (18.5%). Impoverished individuals tend to be single[3] and are less likely to own a home[4] (Haveman & Massaro, 2015).  Among those in poverty, nearly 75% of individuals do not hold a college degree (Haveman & Massaro, 2015).

To address poverty in the Bay Area, the California government has funded multiple programs to assist individuals in need. Below is a table summarizing the description of assistance programs and the number of individuals in Santa Clara County (the largest county population-wise in the Bay Area) receiving benefits:

---

[2] The Bay Area consists of the nine counties that border the San Francisco Bay: San Francisco, San Mateo, Santa Clara, Alameda, Contra Costa, Solano, Napa, Sonoma, and Marin.
[3] Never married (poverty rate of 16.4%), divorced (13.9%), separated (20.8%), or widowed (12.8%).
[4] 55% of the general population owns live in owned homes, while only 21% of individuals in poverty live in owned homes.

| Program | Description | Number of Recipients: 2017 | Portion of Population (1,938,153) |
|---|---|---|---|
| Supplemental Security Income (SSI) | Pays benefits to disabled adults and children who have limited income and resources. | 45,171 | 2.33% |
| California Special Supplemental Nutrition Program for Women, Infants, and Children (WIC) | Provides supplemental foods, nutrition education and referrals to health care, at no cost, to low-income pregnant, breastfeeding and postpartum women, infants, and children up to age 5 who are determined to be at nutritional risk. | 13,652 | 0.70% |
| California Medicaid (Medi-Cal) | Pays for a variety of medical services for children and adults with limited income and resources | 386,781 | 19.96% |
| California CalWORKs | Gives cash aid and services to eligible needy California families. Provides housing, food, utilities, clothing or medical care. | 14,681 | 0.76% |
| CalFresh (Food Stamps) | Provides monthly benefits to assist low-income households in purchasing the food they need to maintain adequate nutritional levels. | 88,471 | 4.56% |
| California Head Start | Promotes the school readiness of children from birth to age five from low-income families by enhancing their cognitive, social, and emotional development. | 2,551 | 0.13% |
| California Low Income Home Energy Assistance Program (LIHEAP) | Provides financial assistance to offset the costs of heating and/or cooling dwellings, and/or have dwellings weatherized to make them more energy efficient. | 7,392 | 0.38% |

*Note.* Data from California Government (n.d.), Santa Clara (2017), "United States" (2017), U.S. Social (2018), Low-Income Home (2017).

Of the individuals receiving benefits from SSI, 55.44% were blind and disabled (U.S. Social, 2018). A total of $28,924,000 was paid out to individuals receiving SSI in 2017 (U.S. Social,

2018). LIHEAP recipients included 1,704 individuals living below 75% of the FPL and 2,606 individuals living from 75% to 100% of the FPL (Low-Income Home, 2017).

From 2011 to 2018, the number of individuals receiving CalWORKs steadily decreased, while the number of individuals receiving Medi-Cal increased (Santa Clara, 2018). There was no trend for CalFRESH recipients, but WIC had a negative trend in participants (Santa Clara, 2018).

## IV. Correlation between Wealth of Community and Poverty

Analyzing differences between high and low wealth communities may provide suggestions for future policy makers on fighting poverty. Using 2016 data from Santa Clara County, we based our comparisons on similar population sizes and contrasting median household income.

### A. Midtown North / Palo Verde / Charleston Gardens vs. Brookwood Terrace

We first compare the community of Midtown North / Palo Verde / Charleston Gardens, located in Palo Alto, CA, with the community of Brookwood Terrace, located in San Jose, CA. While Brookwood Terrace ($43,796) is well below the county average ($93,854), Midtown North / Palo Verde / Charleston Gardens ($156,210) is well above the county average.

| | | Midtown North / Palo Verde / Charleston Gardens | Brookwood Terrace | Santa Clara County |
|---|---|---|---|---|
| **Demographics** | **Population** | 11,692 | 11,774 | 1,781,642 |
| | **Single Parent Households** | 4*% | 13% | 7% |
| | **Households with Children** | 44% | 45% | 39% |
| | **Average Household Size** | 2.69 | 3.82 | 2.90 |
| **Income and Job Opportunities** | **Median Household Income** | $156,210 | $43,796 | $93,854 |
| | **Unemployed (ages ≥ 16 years)** | 7% | 13% | 9% |
| | **Families below 185% FPL** | 4*% | 42% | 16% |
| | **Children (ages 0-17) below 185% FPL** | 2*% | 54% | 25% |
| **Educational Attainment** | **Less than High School (Ages ≥ 25 years)** | 2*% | 40% | 13% |
| | **College Graduate or Higher (Ages ≥ 25 years)** | 82% | 17% | 47% |
| **Food** | **Households Receiving CalFresh Benefits** | 2*% | 14% | 5% |
| **Housing** | **Households Occupied by Renters** | 33% | 57% | 43% |
| | **Households with Gross Rent 30% or more of Household Income** | 36% | 63% | 46% |
| **Safe Communities** | **Average Number of Violent Crimes within 1 Mile** | 1.78 | 28.71 | 16.04 |

| | | Midtown North / Palo Verde / Charleston Gardens | Brookwood Terrace | Santa Clara County |
|---|---|---|---|---|
| Health | Teen Live Births per 1,000 females, ages 15-19 | 2.3 | 58.5 | 19.2 |
| | Adults who are Uninsured (ages 18-64) | 4% to 10% | 19% to 25% | 14% |
| | Life Expectancy | 87.4 | 67.3 | 83.4 |

Note: * indicates estimate is statistically unstable due to a relative standard error of greater than 30%.
*Note.* Data from Midtown North (2016), Brookwood Terrace (2016).

Brookwood Terrace has both a significantly higher percentage of single mothers than Midtown North / Palo Verde / Charleston Gardens and a higher average household size. With an income gap of $112,414 between the two communities, 42% of families and 54% of children are below 185% FPL in Brookwood Terrace while a miniscule amount of families and children are in the wealthier community.

While Brookwood Terrace has less college graduates than Midtown North / Palo Verde / Charleston Gardens, Brookwood Terrace has more crime, more teen births, and more cost-burdened renters. More adults are medically uninsured, and the life expectancy in Brookwood Terrace is roughly 20 years less than in Midtown North / Palo Verde / Charleston Gardens.

### B. Foothills vs. Gilroy - East Side
To analyze trends from the previous comparison, we compare Foothills (located in Los Altos Hills) with East Side of Gilroy. In the following comparison, while the population is slightly smaller, the wealth disparity is greater. Between the two communities, there exists an income gap of $221,776.

| | | Foothills | Gilroy - East Side | Santa Clara County |
|---|---|---|---|---|
| Demographics | Population | 8,537 | 8,459 | 1,781,642 |
| | Single Parent Households | 3*% | 23% | 7% |
| | Households with Children | 31% | 54% | 39% |
| | Average Household Size | 2.79 | 4.04 | 2.90 |
| Income and Job Opportunities | Median Household Income | $258,092 | $36,316 | $93,854 |
| | Unemployed (ages ≥ 16 years) | 5% | 14% | 9% |
| | Families below 185% FPL | 4*% | 61% | 16% |
| | Children (ages 0-17) below 185% FPL | 5*% | 84% | 25% |

| | | Foothills | Gilroy - East Side | Santa Clara County |
|---|---|---|---|---|
| **Educational Attainment** | **Less than High School (Ages $\geq$ 25 years)** | 2*% | 45% | 13% |
| | **College Graduate or Higher (Ages $\geq$ 25 years)** | 84% | 10% | 47% |
| **Food** | **Households Receiving CalFresh Benefits** | 0*% | 26% | 5% |
| **Housing** | **Households Occupied by Renters** | 11% | 73% | 43% |
| | **Households with Gross Rent 30% or more of Household Income** | 27*% | 60% | 46% |
| **Safe Communities** | **Average Number of Violent Crimes within 1 Mile** | 0.32 | 28.78 | 16.04 |
| **Health** | **Teen Live Births per 1,000 females, ages 15-19** | 0.0 | 54.0 | 19.2 |
| | **Adults who are Uninsured (ages 18-64)** | 4% to 10% | 19% to 25% | 14% |
| | **Life Expectancy** | 83.6 | 79.1 | 83.4 |

Note: * indicates estimate is statistically unstable due to a relative standard error of greater than 30%.
*Note.* Data from Foothills Profile (2016), Gilroy - East (2016).

Similar to Brookwood Terrace, Gilroy- East Side has a high percentage of single mothers and large household size. However, Gilroy- East Side has 11% more disparity than Brookwood Terrace regarding single parent households. The vast majority of children (84%) and their families (61%) are living under 185% of the FPL in Gilroy- East Side.

Rated as having the most poverty in Santa Clara County, Gilroy- East Side has 26% of its population receiving CalFresh benefits. Although the entire city of Gilroy consists of only 2.7% of Santa Clara County's population, Gilroy has 8.4% of its population receiving CalWORKS (Gilroy Dispatch Staff, 2015).

73% of households in Gilroy- East Side are occupied by renters, and 60% of those renters are cost-burdened. The median rent in Gilroy was $1,367 over the last five years, which was 48% higher than the nationwide median rent - $920 (Gilroy Dispatch Staff, 2015).

The reason for Gilroy's high poverty levels may stem from the types of jobs in Gilroy. There is a high concentration of temporary and seasonal workers for retail and agriculture sectors. From 2010-2014, retail and agriculture sectors accounted for 19% of Gilroy's workforce (Gilroy Dispatch Staff, 2015). In comparison, only 10.3% of Santa Clara County's workforce was in the retail and agriculture sectors (Gilroy Dispatch Staff, 2015). An individual living in Foothills would not drive to Gilroy to work at a fast-food restaurant. Near Foothills, many tech companies thrive, often offering higher wages for skilled labor.

## C. South Almaden Valley vs. Capital Goss
Lastly, we compare two communities within the city of San Jose. South Almaden Valley and Capital Goss are relatively closer together and fall under the same policies under San Jose.

| | | South Almaden Valley | Capital Goss | Santa Clara County |
|---|---|---|---|---|
| **Demographics** | **Population** | 18,243 | 18,167 | 1,781,642 |
| | **Single Parent Households** | 6% | 13% | 7% |
| | **Households with Children** | 41% | 44% | 39% |
| | **Average Household Size** | 2.96 | 3.77 | 2.90 |
| **Income and Job Opportunities** | **Median Household Income** | $142,923 | $50,130 | $93,854 |
| | **Unemployed (ages ≥ 16 years)** | 7% | 11% | 9% |
| | **Families below 185% FPL** | 4% | 39% | 16% |
| | **Children (ages 0-17) below 185% FPL** | 6*% | 50% | 25% |
| **Educational Attainment** | **Less than High School (Ages ≥ 25 years)** | 4% | 34% | 13% |
| | **College Graduate or Higher (Ages ≥ 25 years)** | 64% | 16% | 47% |
| **Food** | **Households Receiving CalFresh Benefits** | 0*% | 13% | 5% |
| **Housing** | **Households Occupied by Renters** | 15% | 58% | 43% |
| | **Households with Gross Rent 30% or more of Household Income** | 47% | 57% | 46% |
| **Safe Communities** | **Average Number of Violent Crimes within 1 Mile** | 0.98 | 46.57 | 16.04 |
| **Health** | **Teen Live Births per 1,000 females, ages 15-19** | 0.3 | 49.8 | 19.2 |
| | **Adults who are Uninsured (ages 18-64)** | 4% to 10% | 19% to 25% | 14% |
| | **Life Expectancy** | 83.4 | 82.5 | 83.4 |

Note: * indicates estimate is statistically unstable due to a relative standard error of greater than 30%.
*Note.* Data from South Almaden (2016), Capital Goss (2016).

The same trends occur as in the two previous community comparisons; however, life expectancy in South Almaden Valley and Capital Goss are similar- 83.4 and 82.5.

We interviewed Santa Clara County's Policy Analyst of District 1, Heather Wilson[5], to determine policy implications such as resources allocation for San Jose. As South Almaden Valley and Capital Goss both fall under San Jose city policy, Heather says, "we put money where there is a need by analyzing zip codes that have high unemployment and high welfare recipient rates." Heather uses data taken from small areas of each city to determine which areas need the most attention.

---

[5] Heather Wilson covers the cities of Los Gatos, South San Jose, Almaden Valley, Morgan Hill, San Martin, and Gilroy.

Additionally, Heather highlighted, "the county government is designed to help the safety net. For instance, part of our budget needs to go to housing, police, and hospitals. The board of supervisors votes on how to spend the budget money appropriately."

Therefore, San Jose allocates its budget to support Capital Goss, the relatively poorer neighborhood. The data shows that Capital Goss, one of the poorest neighborhoods in San Jose, receives more CalFresh benefits and has higher unemployment than South Almaden Valley.

## V. Solutions: Potential and Past attempts

We review attempts to alleviate the effects of poverty in order to speculate on potential solutions in the future.

Heather provided an example to demonstrate the resources allocation for the city of Saratoga. In Saratoga, many seniors own homes but generate limited income through Social Security. These seniors may spend a significant portion of their limited income on medication. Since Saratoga is a high wealth neighborhood, welfare programs were not implemented - seniors had to drive to neighboring cities to receive welfare benefits. In 2015, Heather collaborated on a project with West Valley Community Services to create a mobile food bank in Saratoga for seniors.

Heather's project helped Saratoga and six neighboring cities. In 2017, West Valley Community Services served 3,893 individuals through distributing 824,644 pounds of food through the food pantry (West Valley Community Services, 2018). $166,544 in financial assistance was provided to support individuals with difficulty paying rent and utility bills (West Valley Community Services, 2018).

Another initiative that has been helping individuals in need is the Silicon Valley Food Rescue. Their website states, "Each year, Santa Clara County is 125 million meals short of feeding its residents in need." To address this issue by rescuing edible food before it enters landfills, Silicon Valley Food Rescue created a food distribution model called "A La Carte," which employs fleets of food trucks to collect surplus food from universities and cafeterias and deliver it directly to individuals in need (Joint Venture Silicon Valley, 2018). As a network of food distribution is created, individuals are encouraged to join the Silicon Valley Food Rescue Association to receive education for food rescuers (Joint Venture Silicon Valley, 2018).

The initiative of Santa Clara County's Supportive Housing System is in place to provide affordable housing to individuals in need. In 2017, 946 permanent supportive housing structures and 503 rapid rehousing structures were built to assist Santa Clara County's vulnerable residents (Santa Clara County, 2017). One project that the county is currently working on is the Leigh Avenue Senior Apartments project in San Jose, which is projected to finish in March 2020 with 64 total units (Santa Clara County, 2017).

Some recommendations for future planning include the establishment of a regional working group and the hiring of a county-level Food Rescue Coordinator (Katz & Rivero, 2015). Multiple food sharing networks also would assist the rescue and transportation of food in regional areas. Katz and Rivero (2015) estimate that "as much as 26 million meals could be served and 25 million pounds of greenhouse gases could be reduced" if food rescue were prioritized instead of only composting. One last suggestion is to support legislation that helps individuals support others who are in need; AB 234, for instance, is a policy that, if approved, would help residents donate food from their homes without special permits.

Through the culmination of these efforts, and through proper future policy planning, we can hope for a world devoid of homelessness in the near future.